# Vapor-to-glass preparation of biaxially aligned organic semiconductors

Jianzhu Ju *[,a], Debaditya Chatterjee [b], Paul M. Voyles [b], Harald Bock [c], Mark D. Ediger *[,a]

[a] Department of Chemistry, University of Wisconsin-Madison, Madison, Wisconsin 53706, United States
[b] Department of Materials Science and Engineering, University of Wisconsin-Madison, Madison, WI 53706, USA
[c] Centre de Recherche Paul Pascal, CNRS & Université de Bordeaux, 33600 Pessac, France

* jianzhu.ju@outlook.com
* ediger@chem.wisc.edu

## Abstract

Physical vapor deposition (PVD) provides a route to prepare highly stable and anisotropic organic glasses which are utilized in multi-layer structures such as organic light emitting devices (OLEDs). While previous work has demonstrated that anisotropic glasses with uniaxial symmetry can be prepared by PVD, here we prepare *biaxially* aligned glasses in which molecular orientation has a preferred in-plane direction. With the collective effect of the surface equilibration mechanism and template growth on an aligned substrate, macroscopic biaxial alignment is achieved in depositions as much as 180 K below the clearing point $T_{LC-iso}$ (and 50 K below the glass transition temperature $T_g$ ) with single-component disk-like (phenanthroperylene ester) and rod-like (itraconazole) mesogens. The preparation of biaxially aligned organic semiconductors adds a new dimension of structural control for vapor-deposited glasses, and may enable polarized emission and in-plane control of charge mobility.

Physical vapor deposition (PVD) [1, 2] provides a unique approach to preparing glassy materials with exceptional kinetic stability [1, 3-5] and anisotropic order [6, 7]. Molecular orientation can be tuned over a wide range by controlling the substrate temperature and deposition rate [3, 8, 9]. PVD has an additional advantage for the production of multilayer structures, where the solid-state preparation of individual layers below the glass transition temperature ($T_g$) avoids the dissolving/melting of previously deposited layers [10-12]. PVD is widely applied in the production of organic light-emitting diodes (OLEDs). Uniaxially anisotropic glassy structures have proven advantageous in OLEDs, by increasing charge mobility [13, 14] and outcoupling efficiency [15-17]. In addition, the high kinetic stability of PVD glasses has been shown to increase device lifetime [18].

Previous work [3, 6, 8, 19] has described how uniaxial anisotropic molecular packing and high kinetics stability is generated in PVD glasses through a surface equilibration mechanism. Molecules at the surface of organic glasses possess enhanced mobility[1, 9, 20, 21] and this mobility allows partial equilibration during deposition. Since this efficient equilibration occurs below the conventional $T_g$, it can lead to glass that have low enthalpy, high density, and high kinetic stability. Equilibration during deposition also produces packing arrangements that are generally anisotropic, as the equilibration takes place in an anisotropic environment (the free surface). After being buried by subsequent deposition, molecules lose their high mobility and are locked in packing arrangements preferred at the surface [6, 22]. Deposition rate and substrate temperature control the final anisotropic structure of the glass because these parameters control the extent to which equilibration occurs at the free surface during deposition [8, 23, 24]. As the result of this free surface effect, the anisotropy of PVD glasses exhibits *uniaxial* symmetry with respect to the surface normal. For example, rod-like molecules in PVD prefer vertical orientation at high substrate temperature $T_{sub}$ and horizontal orientation at low $T_{sub}$: in both cases, macroscopic samples are isotropic in-plane [8, 9, 25].

The existence of uniaxially anisotropic glasses naturally leads to the question of whether an even greater degree of anisotropic order can be prepared via PVD: Can we prepare *biaxially* anisotropic glasses? Molecules in a biaxial PVD film would have a preferred in-plane orientation, in addition to having a preferred orientation with respect to the surface normal. Beyond the fundamental challenge, we expect that PVD preparation of biaxial glassy anisotropy, can lead to new material properties that will enable new devices. A potential route to obtain biaxial glassy structures is to combine the surface equilibration mechanism with an additional physical mechanism to control alignment along different in-plane directions. Template growth is a method often applied to prepare crystalline nanostructures with well-controlled morphology and structural properties [26-28]. It takes advantages of an epitaxial interaction between the substrate and the molecules, which is not present for typical PVD depositions [29]. To use template growth as an additional control mechanism for anisotropy in organic materials, we propose in this work to use PVD to deposit glass-forming mesogens [30] onto an in-plane alignment substrate. The interactions of mesogens (i.e. liquid crystal molecules) with substrates has been widely studied in their fluid (i.e. non-glassy) liquid crystalline mesophase [31-34], where typically a thick mesogen layer over a few microns can be macroscopically aligned between pre-orientated parallel substrates [35-38]. As we describe below, PVD deposition onto an alignment substrate is distinct from this, as a thin film is formed directly in the solid state, which is a key advantage. It is well established that mobility during PVD of glassy materials at temperatures below $T_g$ [9, 25] is limited to the free surface (within a few nanometers) [3, 8]. We envision that this mobility allows the initially deposited molecules to be aligned by the substrate. Each layer of deposited mesogens should then behave as a template layer for alignment of the following mesogens during PVD, with the entire film growth occurring below $T_g$.

Here we show that PVD can combine the surface equilibration mechanism with template growth in order to produce thin films of biaxially aligned organic semiconductors. We quantify the degree of biaxial order with X-ray scattering and polarized microscopy. We show that the successful control of biaxial order for the entire thickness of a deposited glass requires that the deposited mesogens have the capability to: 1) interact with the in-plane alignment substrate during initial deposition of mesogens; 2) maintain the in-plane alignment in subsequent deposition, i.e. template growth in the glassy state; 3) retain control of molecular orientation with respect to the surface normal, as controlled by the surface equilibration mechanism. The preparation of biaxially aligned organic semiconductors may provide new functionality for organic electronics, including polarized emission and in-plane control of charge mobility.

## Experimental section

As the alignment substrate, polycarbonate ($M_w$ = 64000 g/mol) was dissolved by stirring overnight in dichloromethane at a concentration of 2 wt%. Spin coating was performed with rate of 6000 rpm for 1 min. Then the wafers were kept in the fume hood for more than 5 hours to evaporate the solvent and form a better attachment wtih the PC layer. The thickness of the PC layer is around 130 nm, measured by an M-2000U spectroscopic ellipsometer from J.A. Woollam Co., Inc.. The bottom surface of a 500 g weight was covered with velvet cloth (Buehler MasterTex polishing cloth) to generate the oriented polymer surface. To apply reproducible rubbing, the weight was slid horizontally along the wafer unidirectionally, for around 1 s per pass. The procedure provides a constant pressure of 1000 Pa and rubbing speed of ~2.5 cm/s. Rubbing results in grooves at an interval of 500 nm to 1 $\mu$m, with a depth of a few nanometers (Fig. S2(c)), observable by atomic force microscopy (in Figure S2 in **Supporting information**).

Physical vapor deposition (PVD) process has been described in detail elsewhere [7]. PVD was performed in a vacuum chamber with pressure around $10^{-7}$ torr. The source was heated at the bottom of the chamber to produce vapor and the deposition rate is monitored with a quartz crystal microbalance (QCM), with precision of 0.01 nm/s. The synthesis of phenanthroperylene ester has been described in detail in ref. [39]. Itraconazole was purchased from Sigma-Aldrich and used as received. Substrates were adhered with Apiezon H grease to temperature-controlled copper blocks to achieve different $T_{sub}$.

The 4D-STEM measurements have been described in detail elsewhere [40]. Phenanthroperylene ester was deposited on a grid with an ultrathin carbon layer (5-6 nm) on top (CF400-CU, Electron Microscopy Sciences). Orientation mapping was performed with a 200 kV Thermo Fisher FEI Titan microscope in $\mu$P EFSTEM mode. Diffraction patterns were collected using a Direct Electron Celeritas detector with 256×256 readout area and resolution of $q$ = 0.314 $nm^{-1}$. Spatially-resolved diffraction patterns were azimuthally integrated between $q$ = 17 to 20.7 $nm^{-1}$ and then fitted to obtain the orientation maps. The domain size was estimated by orientation correlation $C$, as shown in **Supporting information**.

Grazing incidence wide angle X-ray scattering (GIWAXS) for static measurements were performed in Stanford Synchrotron Radiation Light (SSRL) source on Beamline 11–3, with wavelength of 0.0976 nm. The distance between the sample and detector was 317 mm, calibrated with $LaB_6$. The measurements were performed in helium at room temperature around 300 K, with exposure time of 60 s per image. The beam size is 0.15 mm × 0.15 mm, and sample is reciprocated in 4 mm range perpendicularly to incidence direction. With grazing-incidence angle of 0.2 °, the detected diffraction is averaged in a narrow region with width of 4 mm across the whole sample (over 25 mm), i.e., a macroscopic region. Incidence angle is 0.2 ° for all samples and all raw diffraction patterns were "$\chi$ corrected" which results in the missing wedge [41]. The diffraction from PC was removed as background for all patterns shown in the paper.

Polarized reflected light microscopy was performed with Olympus BX51 in reflection mode. With polarizer and analyser perpendicular to each other, the maximum and minimum reflected intensity (transmitted twice through the film) can be obtained when the polarizer is 45 ° and 0 ° to the rubbing direction, respectively.

**Molecular orientation can be templated during PVD**

A critical feature of our proposed alignment mechanism is the capability for templating of alignment during deposition into the glassy state, so we first demonstrate that this is physically possible. A discotic mesogen, phenanthroperylene ester [39] (phen-ester for simplicity), is chosen as the model mesogen. For deposition at high substrate temperature $T_{sub}$ just below $T_g$ = 392 K, phen-ester forms a hexagonal columnar mesophase, with edge-on molecular orientation (Figure 1(a)) [7, 8]. Previous work used four-dimensional scanning transmission electron microscopy (4D-STEM) to show that the in-plane molecular orientation forms nanoscale domains 10-20 nm across [40]. Based on that capability, here we demonstrate that in-plane mesogen alignment can be transferred through many layers of subsequently deposited material.

Figure 1(b) shows that, phen-ester is deposited onto ultrathin TEM grids with $T_{sub}$ = 392 K and a deposition of rate 0.03 nm/s, to produce a 20 nm thick film. Orientation mapping of $\pi - \pi$ stacking can be obtained by spatially resolving diffraction patterns. Domains with uniform orientation can be detected. The characteristic domain size is $13 \pm 1$ nm, as calculated by the method described in ref. [40] and shown in the **Supporting information**. We emphasize that the observed domains are three-dimensional and extend throughout the film's thickness [40]. When we use the 392 K deposited phen-ester film (a separate one than that in Figure 1(b)) as a substrate to prepare a secondary 20 nm thick layer at $T_{sub}$ = 370 K, the bilayer film shows a similar domain size of 15 $\pm$ 1 nm (Figure 1(c)). In contrast, deposition of a 40 nm thick film at 370 K produces smaller domains of 8 $\pm$ 1 nm (Figure 1(d) and also results in ref [40]).

The results in Figure 1(c) clearly show that the bottom 392 K layer with large domains templates the local orientation of the top layer at $T_{sub}$ = 370 K. The successful templating observed here in the robust solid mesophase ($T_g$ - 20 K) is a surface effect enabled by high surface mobility, which distinguishes it from templating in a fluid mesophase. For bulk liquid crystals, the high mobility required for global self-assembly can only be achieved at temperatures close to clearing temperature $T_{LC-iso}$ [42, 43] (520 K for phen-ester [44]).

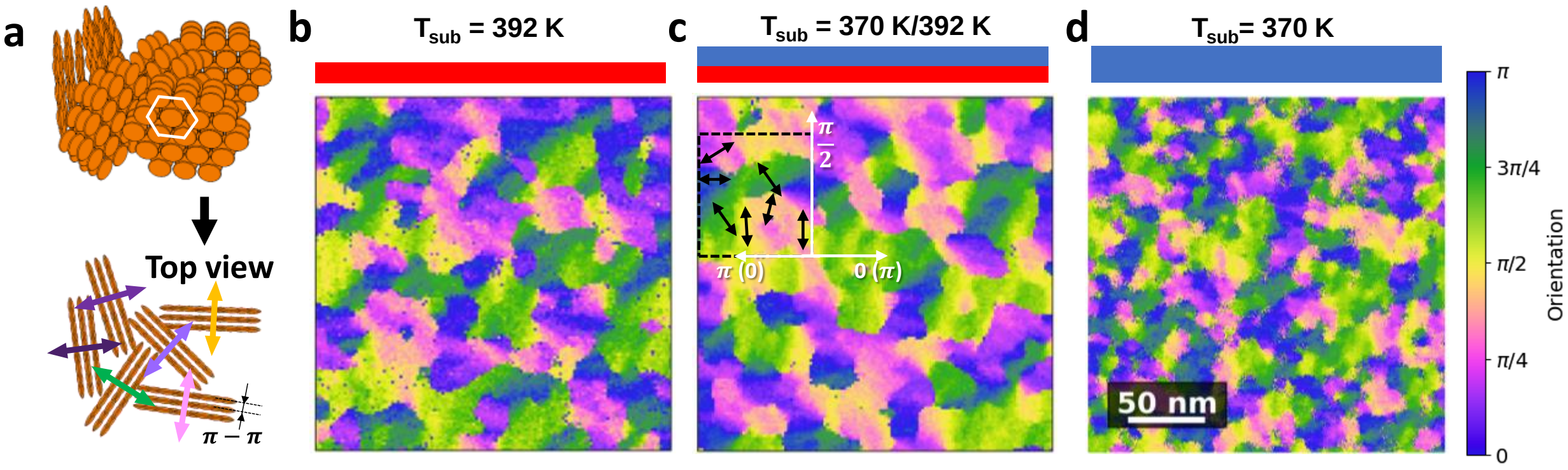

*Figure 1 (a) Schematic drawing of phen-ester hexagonal columnar phase, as produced by PVD on an isotropic substrate at high $T_{sub}$. The top view image corresponds to the observation by 4D-STEM where the orientation of $\pi - \pi$ stacking can be mapped with spatial resolution of 2 nm. Orientation maps are plotted in a field of 252 nm for PVD glasses prepared at $T_{sub}$ of (b) 392 K, (c) 370 K on 392 K (bilayer film) and (d) 370 K. A small region in Figure 1(c) indicates the orientation of individual domains with black arrows.*

## Macroscopic alignment of mesogen orientation

Having verified that orientation templating occurs locally during PVD, we attempted to prepare thin films with macroscopic in-plane orientation by depositing onto an alignment substrate, as described in **Materials and Methods.** The alignment substrate was prepared by spin-coating polycarbonate (PC) with 130 nm thickness onto a 25 mm diameter silicon wafer, followed by mechanical rubbing. 250 nm of Phen-ester was deposited onto the PC layer at different $T_{sub}$, at a deposition rate 0.06 nm/s. Mechanical rubbing results in grooved surface with thickness fluctuation in a scale of a few nanometers (as seen in AFM measurements in Figure S2 in **supporting information**), which is much smaller than the deposited layer, so that it can be safely neglected. The in-plane anisotropy of the deposited phen-ester was measured with grazing-incidence wide-angle X-ray scattering (GIWAXS) by rotating the sample.

GIWAXS patterns verifying macroscopic in-plane mesogen orientation are shown in Figure 2(a). The diffraction from PC was removed as background for all patterns shown in the paper. The in-plane angle between the X-ray beam and the rubbing direction is denoted as $\psi_{view}$, where $\psi_{view}$ = 0 ° when the beam is parallel to the rubbing direction (Figure 2(b)). At $T_{sub}$ = 392 K, for both $\psi_{view}$ = 0° and 90°, the hexagonal columnar phase can be detected with spot-like patterns at $\chi$ of 0 ° and 60 ° ($\chi$ = 0° is defined to be along $q_z$ [8]), at $q$ around 4 nm$^{-1}$; we designate the peak at $\chi$ = 60 ° as Peak H. $\pi$-$\pi$ stacking gives rise to a peak at $q$ =18 nm$^{-1}$ along $q_{xy}$ at small $q_z$, and we designate this as Peak $\pi$. Considering both patterns, we confirm that the hexagonal columnar phase with edge-on molecule orientation is formed, similar to that formed on an isotropic substrate at high $T_{sub}$ [8], as also seen in Figure S3. Remarkably, Peak H is much brighter at $\psi_{view}$ = 0° and Peak $\pi$ is much brighter at $\psi_{view}$ = 90°. From this result, the anisotropic packing of the hexagonal columnar phase can be understood, with most of the columns propagating along the rubbing direction. [For perfect alignment (schematically shown in Figure 2(b)), Peak $H$ would only be observed at $\psi_{view}$ = 0 ° and peak-$\pi$ only at $\psi_{view}$ = 90 °.] The anisotropy is also observed for deposition at 370 K, as shown in Figure 2(a). At $T_{sub}$ below 370 K, the hexagonal columnar packing is barely identifiable, and the diffraction patterns are identical at all $\psi_{view}$ (Figure S4). In summary of the GIWAXS patterns, a macroscopically biaxial anisotropy can be obtained by PVD on an alignment substrate well below $T_{LC-iso}$ and even in glass state below $T_g$. Importantly, we observe a highly anisotropic structure in film with thickness of 250 nm. This extends the results of the 4D-STEM measurements (Figure 1(b)) which demonstrated effective templating through the thickness of 40 nm films.

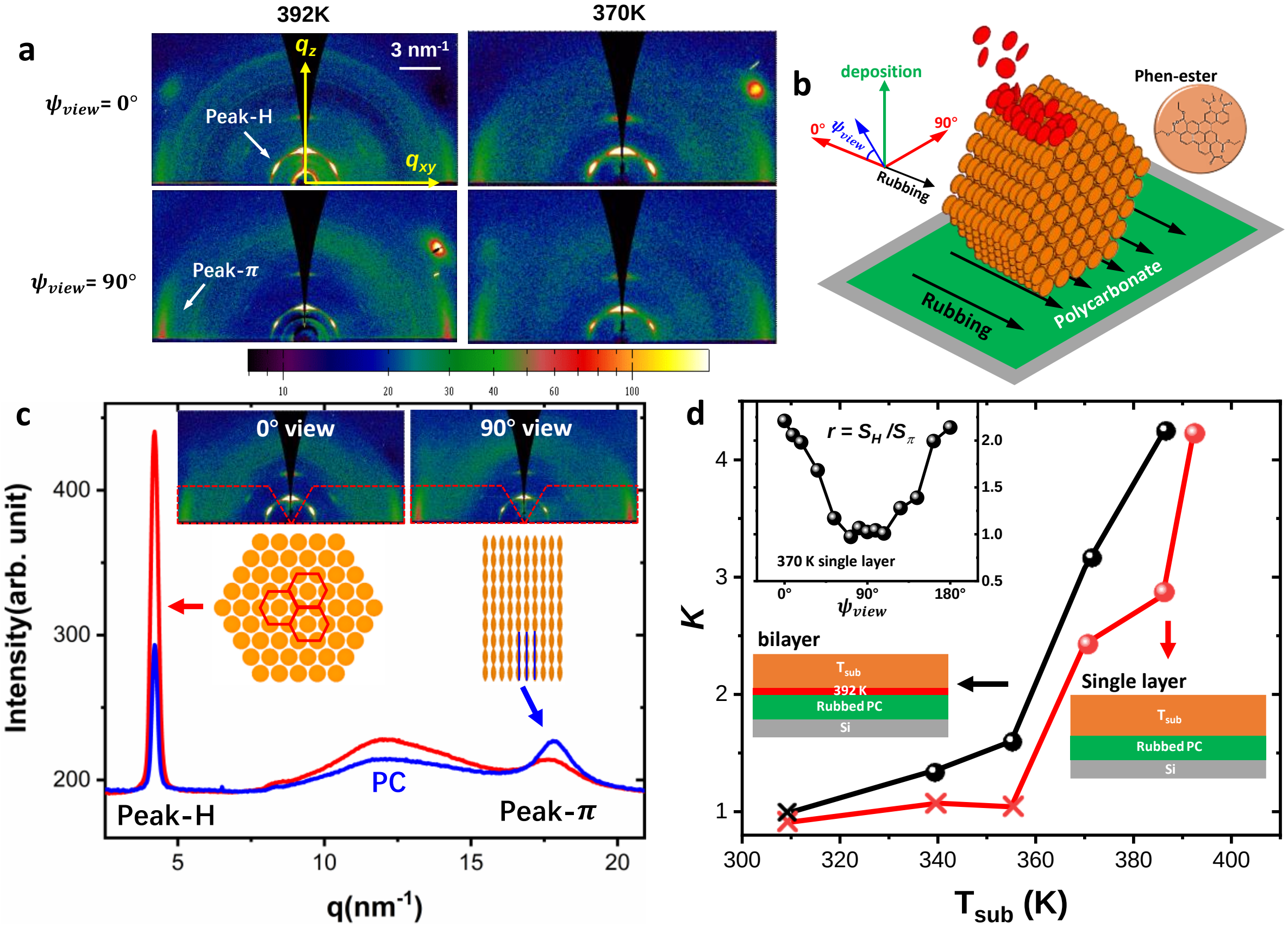

*Figure 2 (a) GIWAXS patterns at different $T_{sub}$ with $\psi_{view}$ = 0 ° and 90 °. (b) Schematic of biaxial alignment of phen-ester. (c) 1D intensity as the function of q for 386 K bilayer (integrated in the region marked in the inset)). The curves are normalized to the intensity at q = 30 nm-1. Inset: GIWAXS patterns and schematic drawing at 0 ° and 90 ° view. (d) Orientation parameter K as the function of $T_{sub}$ for single layers (red) and bilayers (black). Inset curve: Peak area ratio $S_H/S_\pi$ as a function of $\psi_{view}$ for 370 K single layer.*

Inspired by the 4D-STEM observation in Figure 1, we tested whether the macroscopic alignment efficiency can be further improved by introducing a thin template layer deposited at higher $T_{sub}$. The deposition of the bilayer films follows two steps: first, a 30 nm layer of phen-ester is deposited onto the PC alignment substrate at $T_{sub} = T_g$ = 392 K, as a templating layer; then a 220 nm layer is deposited onto the templating layer at different $T_{sub}$. To quantitatively characterize the macroscopic alignment, the averaged intensity as a function of $q$ is calculated. [Intensity is averaged in the region indicated in Figure 2(c)): with $q_z$ < 3.7 nm$^{-1}$ and excluding the region containing the vertical peak at $q$ around 4 nm$^{-1}$ ($-30° < \chi < 30°$), which is contributed from all $\psi_{view}$.] 1D intensity curves for the bilayer at $T_{sub}$ = 386 K (on top of thin 392 K layer) at different $\psi_{view}$ are shown in Figure 2(c). Significant and *opposite* changes in the intensity of peak-H and peak-$\pi$ can be observed. Peak areas $S_H$ and $S_\pi$ can be fitted from 1D intensities (details shown in Figure S5). The wide peak at $q \sim 13$ nm$^{-1}$ is related to PC and can be well-separated from $S_H$ and $S_\pi$. The ratio $r(\psi_{view}) = S_H/S_\pi$ is a characteristic parameter for in-plane anisotropy, with the two peaks having opposite dependences on $\psi_{view}$. Higher $r$ means a larger number of hexagonal columns propagating in the direction of the X-ray beam. $r(\psi_{view})$ for the single layer deposition at $T_{sub}$ = 370 K is plotted in the inset of Figure 2(d), where more than a two-fold change can be observed. The macroscopic alignment is quantified by an anisotropy parameter $K$, which increases with in-plane anisotropy:

$$K = \frac{r(0°)}{r(90°)} = \frac{S_{H0}S_{\pi 90}}{S_{\pi 0}S_{H90}}$$

Figure 2(d) compares the anisotropy parameter $K(T_{sub})$ in bilayers and single layers. In both series, $K$ increases with $T_{sub}$ from 1 (isotropic in-plane) at low temperature to 4.5 for $T_{sub} \sim T_g$. This is expected since the mesogens have higher mobility for better surface equilibration and template interaction when deposited at high $T_{sub}$ [8, 9]. Comparing the single layers and bilayers, even though 88% of the phen-ester is deposited at same $T_{sub}$, we find that $K$ for the bilayer is significantly larger than for the single layers. A more common orientation parameter $S_{2D}$ can be calculated from multi- $\psi_{view}$ measurements [45] as shown in Figure S6 in **Supporting information**. For the sample with highest orientation, the bilayer at $T_{sub}$ = 386 K, we estimate $S_{2D}$ = 0.37, showing substantial anisotropy.

The superiority of bilayers over single layers is notable, given that all the observed in-plane anisotropy is ultimately derived from the same aligned PC substrate. We interpret this to mean that the substantial alignment shown in Figure 2(d) is not limited by the choice of mesogen or by the PVD process, and that even greater macroscopic anisotropy could be achieved with a substrate that more effectively aligns (in comparison to PC). AFM measurements show that the 392 K template layer is smoother than the PC layer itself (Figure S2 in **Supporting information**), emphasizing that the alignment efficiency observed in the bilayer samples is dominated by intramolecular interaction at the surface. Remarkably, biaxial anisotropy can still be generated at 180 K below $T_{LC-iso}$, in contrast to conventional liquid crystal alignment.

Additional experiments show that conventional alignment in the fluid mesophase is impossible with this mesogen and substrate. We vapor-deposited phen-ester films on rubbed PC below $T_g$, and then annealed at around 420 K while acquiring GIWAXS data (Figure S9 in **Supporting information**). No alignment or rearrangement of the megogens could be detected, as expected for annealing far below the clearing temperature (520 K). Higher annealing temperatures are not feasible, as annealing above the PC $T_g$ (420 K [46, 47]) destroys the alignment in the PC substrate. This illustrates a key advantage of our PVD approach. Even substrates with relatively low $T_g$ values can be used for alignment. Additionally, the present route produces alignment on films as thin as 20 nm without dewetting (Figure 1). This cannot be achieved in the fluid mesophase because the high temperatures required for alignment also lead to dewetting.

**Biaxial alignment by organic semiconductor substrate**

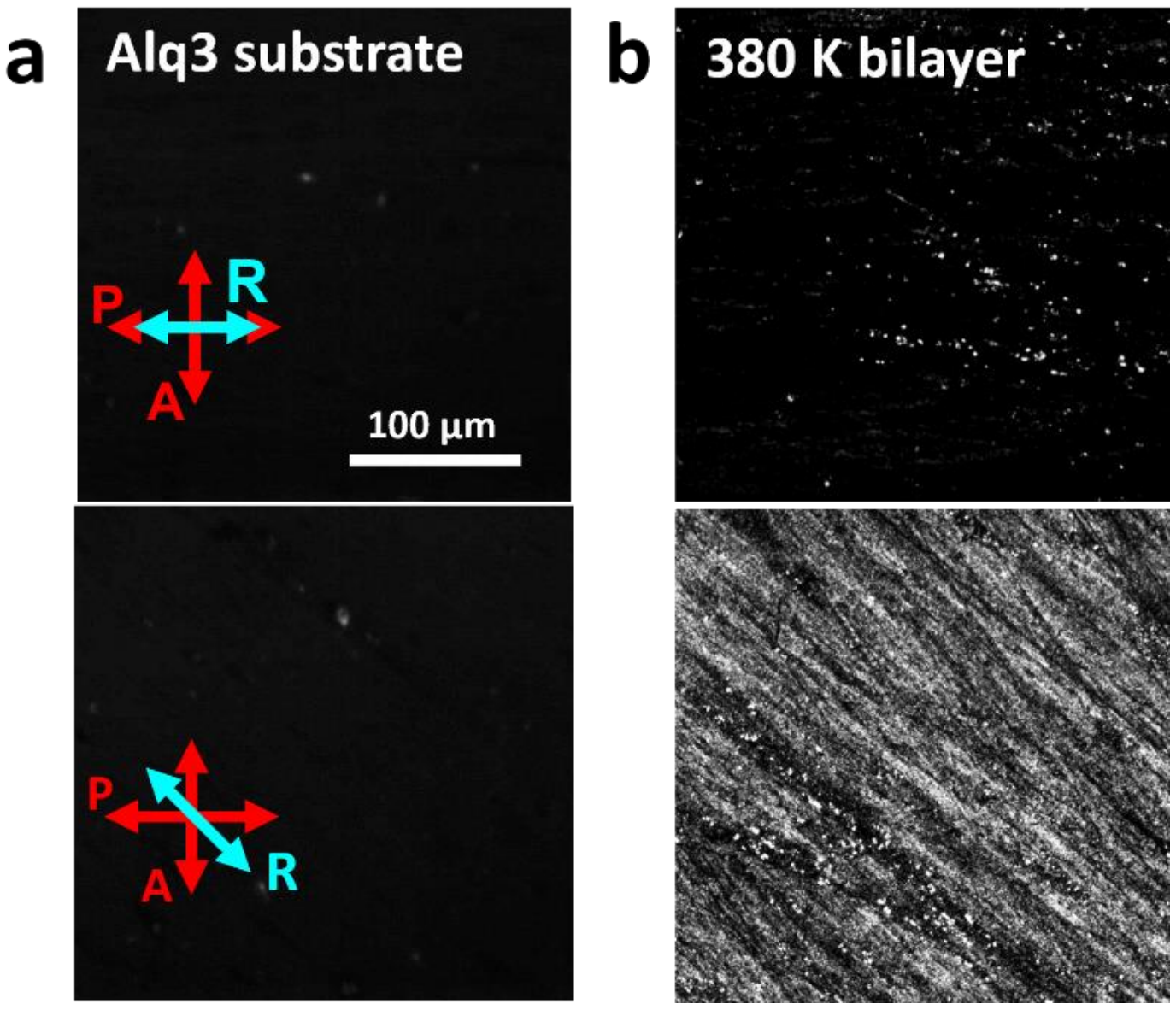

*Figure 3 Optical microscopy images utilizing crossed polarizers with the rubbing angle at 0 ° (top) and 45 ° (bottom) for (a) Alq3 substrate and (b) 380 K phen-ester bilayer. The rubbed Alq3 substrate is not birefringent, while the phen-ester overlayer is strongly birefringent (and thus aligned).*

To further test the feasibility of coupling biaxial alignment by PVD with multilayer device manufacturing, we show that alignment can also be achieved using an organic semiconductor as the substrate. A 300 nm layer of $Alq_3$ (tris(8-hydroxyquinoline) aluminum)[48] was vapor-deposited (at 340 K with rate of 0.2 nm/s) onto Si wafer and then rubbed with the process discussed above. On top of this, we vapor-deposited a 250 nm phen-ester bilayer (20 nm at 392 K, followed by 230 nm at 380 K). Optical anisotropy is quantified by the degree of polarization $D$, defined as

$$D = \frac{I_{max} - I_{min}}{I_{max} + I_{min}} ,$$

where $I_{max}$ and $I_{min}$ can be obtained when the crossed polarizers are 45 ° and 0 ° to the rubbing direction, respectively. As shown in Figure 3(a), the rubbed Alq3 substrate exhibits no birefringence in an optical microscope (presumably as a result of the molecule's nearly spherical symmetry [48]). On the other hand, in Figure 3(b), strong birefringence can be observed in 380 K bilayer, with $D$ of 70%. The large birefringence observed when the phen-ester bilayer is present can be completely attributed to in-plane orientation of phen-ester molecules, as expected for a strongly aligned sample. GIWAXS measurements of the 380 K bilayer on $Alq_3$ (Figure S7) confirm this interpretation, indicating that K = 2.7. All these observations indicate that biaxial alignment of semiconductor mesogens by PVD can be achieved with an underlying organic semiconductor substrate, and also shows that PVD can effectively prepare well-aligned nanofilms with substantial optical anisotropy.

**Biaxial alignment of rod-like mesogens**

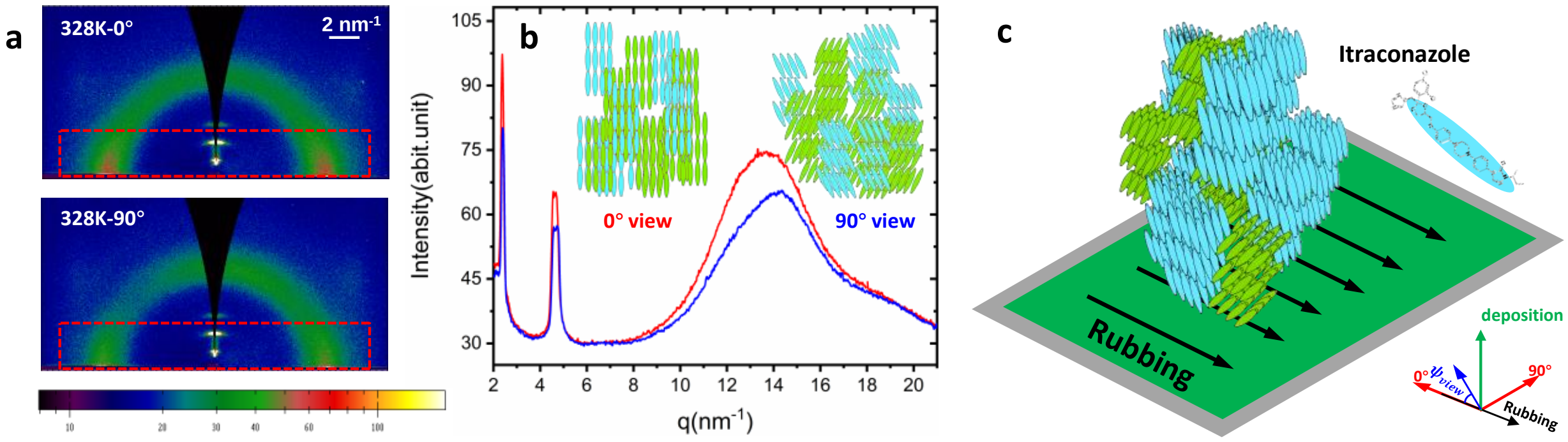

*Figure 4 (a) GIWAXS patterns of a rod-like mesogen, itraconazole, at $\psi_{view}$ =0 ° and 90 ° with $T_{sub}$ = $T_g$ = 328 K. (b) 1D intensity averaged in the region shown in (a). Inset: schematic 0 ° view and 90 ° corresponding to GIWAXS patterns. (c) Schematic drawing of the biaxial alignment of itraconazole.*

The preparation route described above should be applicable for any mesogens with the ability to generate biaxial packing. Here we show an additional example of solid-state biaxial alignment, with a rod-like mesogen, itraconazole. As reported in previous work [9, 49], at high $T_{sub}$, deposition of itraconazole on an isotropic substrate generates a glass with uniaxial anisotropy, with mesogens tilted at an average angle around 60 ° from the substrate [49]. To achieve biaxial alignment, we deposited itraconazole at $T_{sub}$ = $T_g$ = 328 K ($T_{LC-iso}$ = 363 K [50]), on rubbed PC. GIWAXS experiments with $\psi_{view}$ = 0 ° and 90 ° (Figure 4(a)) show a layered structure (diffraction peaks at $\chi = 0°$ with first peak at $q_z$ around 2 nm$^{-1}$) with overall vertical mesogen orientation ($\pi$-$\pi$ stacking peaks at $\chi$ = 90° and $q_{xy}$ around 14 nm$^{-1}$) can be observed. The horizontal peak is stronger at $\psi_{view}$ =0 °, indicating in-plane anisotropy. To avoid scattering from PC (see **Supporting information**), an averaging region containing mainly the in-plane peaks (indicated in Figure 4(a)) is selected for quantitative comparison, with results shown in Figure 4(b). The in-plane anisotropy results from the molecules preferentially tilting along the rubbing direction [49], as schematically shown in Figure 4(c). We expect that a perfectly aligned structure, at $\psi_{view}$ = 0 °, would show no indication of tilting, while at $\psi_{view}$ = 90 ° tilted orientation would be detected (inset in Figure 4(b)).

## Concluding remarks

In this work, we report substantial progress towards the manufacture of *biaxially* aligned glasses by PVD. A glassy solid with in-plane alignment is directly synthesized at low temperature without utilizing equilibrium liquid crystalline phases. In this process, mesogens also exhibit out-of-plane orientation, which is similar to when they are deposited on isotropic substrates. In this way, the out-of-plane and in-plane orientation are *independently* governed by the free surface and the alignment substrate, respectively. A key parameter, $T_{sub}$, tunes the extent of equilibration for both out-of-plane and in-plane orientation.

The successful preparation of biaxially-aligned glass makes use of the stimuli-responsiveness of mesogenic vapors to an oriented substrate. During PVD, this can occur below $T_g$ as a result of high molecular mobility at the free surface of the growing film, and transfer of in-plane alignment from layer to layer via templating. This solid-state low temperature processing can achieve substantial alignment at up to 180 K below $T_{LC-iso}$, thus avoiding high temperature preparation. We have also shown that a thin glassy layer of the mesogen can be used as a template to better align the following layers. Such templating layers could be separately programable (via $T_{\mathrm{sub}}$) during a continuous deposition, providing further process control to increase or decrease in-plane alignment.

At the current state of development, the preparation of a biaxially aligned glass depends on a pre-processed alignment substrate that cannot be adjusted *in situ*. More to this, mechanical rubbing unavoidably increases surface roughness, which is unfavored in electronic devices with ultrathin layers. As a next step, strategies that allow programmed alignment inside a deposition chamber without mechanical rubbing will be explored. Strategies borrowed from bulk liquid crystals may be successful [51], such as coupling electric and magnetic fields into the deposition system, or preparing an alignment layer with an anisotropic surface morphology that can be prepared by glancing angle vapor deposition [52-56] .

What are the prospects for using PVD to produce biaxial glasses of non-mesogens? In qualitative terms, glasses are not very good at propagating structure away from an interface, because there are so many available packing arrangements. The packing arrangements of mesogens are sufficiently limited that structure can be successfully propagated from the substrate, as shown in this work. It is an open question whether PVD of non-mesogens can be used to prepared biaxial glasses.

## Supplemental Material

Supplemental Material can be found in: …. Additional informations are provided:

1. Data processing of 4D-STEM data;
2. AFM images showing surface morphology of the rubbed substrate;
3. GIWAXS analysis detail and additional GIWAXS results;

## Acknowledgements

Primary funding for the work was provided by the Wisconsin MRSEC (NSF DMR-1720415). The authors thank Xiaodan Gu and Guorong Ma for kindly providing advice and assistance with in-situ GIWAXS measurements, and Dean DeLongchamp for useful discussions. Use of the Stanford Synchrotron Radiation Lightsource, SLAC National Accelerator Laboratory was supported by the U.S. Department of Energy, Office of Science, Office of Basic Energy Sciences under Contract No. DE-AC02-76SF00515.

## Data Availability

ASCII and tif. format images in all the presented figures have been uploaded in Zenodo (DOI: 10.5281/zenodo.7747240).

# Supporting information

## 1. Domain size quantification from 4D-STEM orientation maps

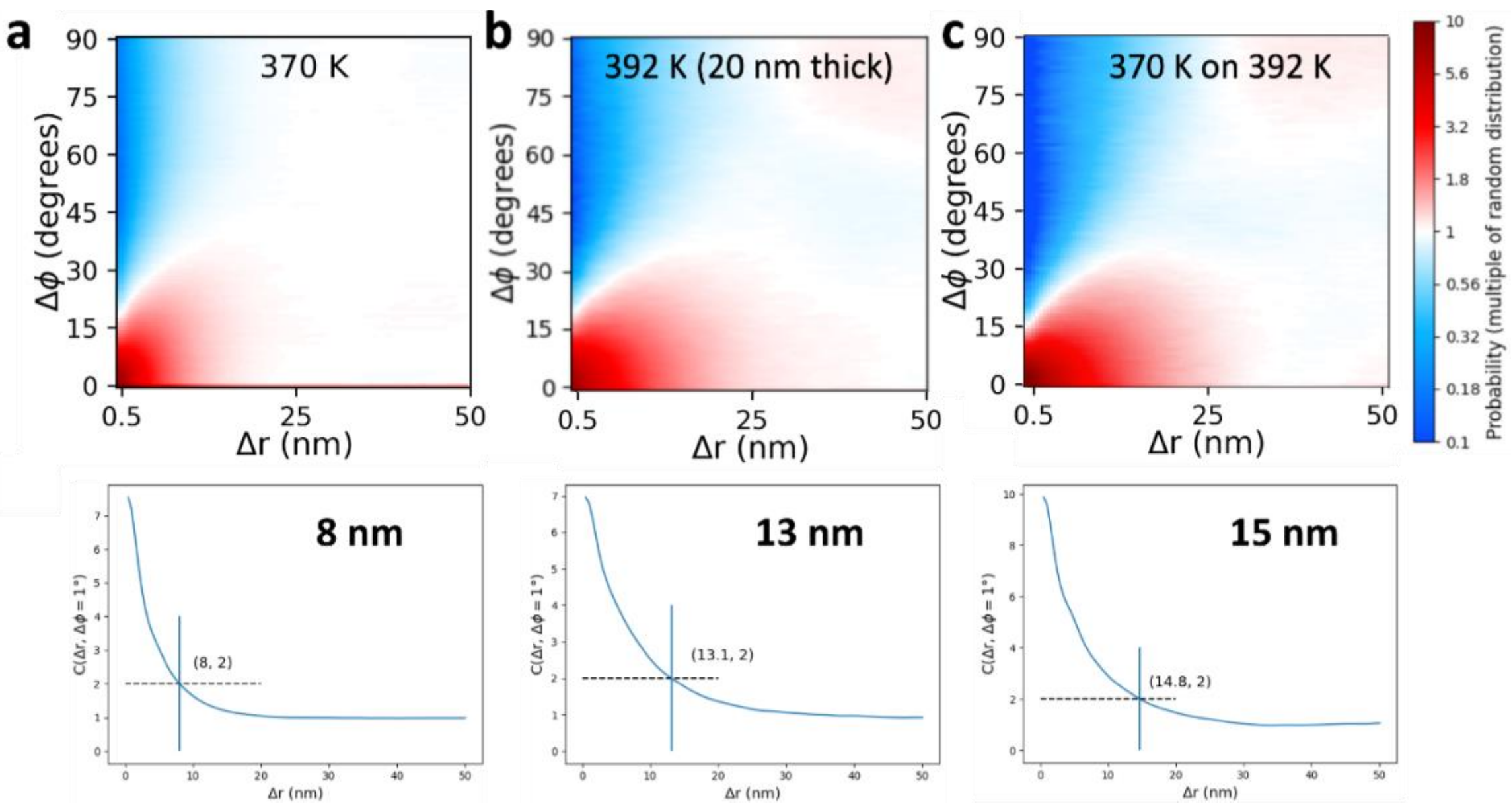

*Figure S1 The processing of 4D-STEM data for (a) 370 K, (b) 392 K 20 nm and (c) 370 K on 392 K (bilayer film). Top: the orientation correlation $C$ as the function of distance $\Delta r$ and misorientation $\Delta\phi$; bottom: $C$ at $\Delta\phi$ = 1 °, as the function of $\Delta r$. Characteristic domain size is defined as the $\Delta r$ value for $C$ = 2.*

From 4D-STEM orientation maps shown in Figure 1, the characteristic domain size can be calculated, defined as the $\Delta r$ value at $\Delta\phi$ = 1 ° and $C$ = 2, as shown in Figure S1, where $\Delta r$ is the correlation length. The detail of the analysis was described in ref. [1]. The uncertainties in the domain size is estimated as ± 1 nm based on the spread in similar data.

## 2. Comparison between rubbed PC and template layer deposited at 392 K

Figure S2 shows the atomic force microscopy (AFM) images of the surface of polycarbonate layer after rubbing (Figure S2(a)) and phen-ester template layer deposited at 392 K (Figure S2(b)). It can be observed that the phen-ester template surface is smoother.

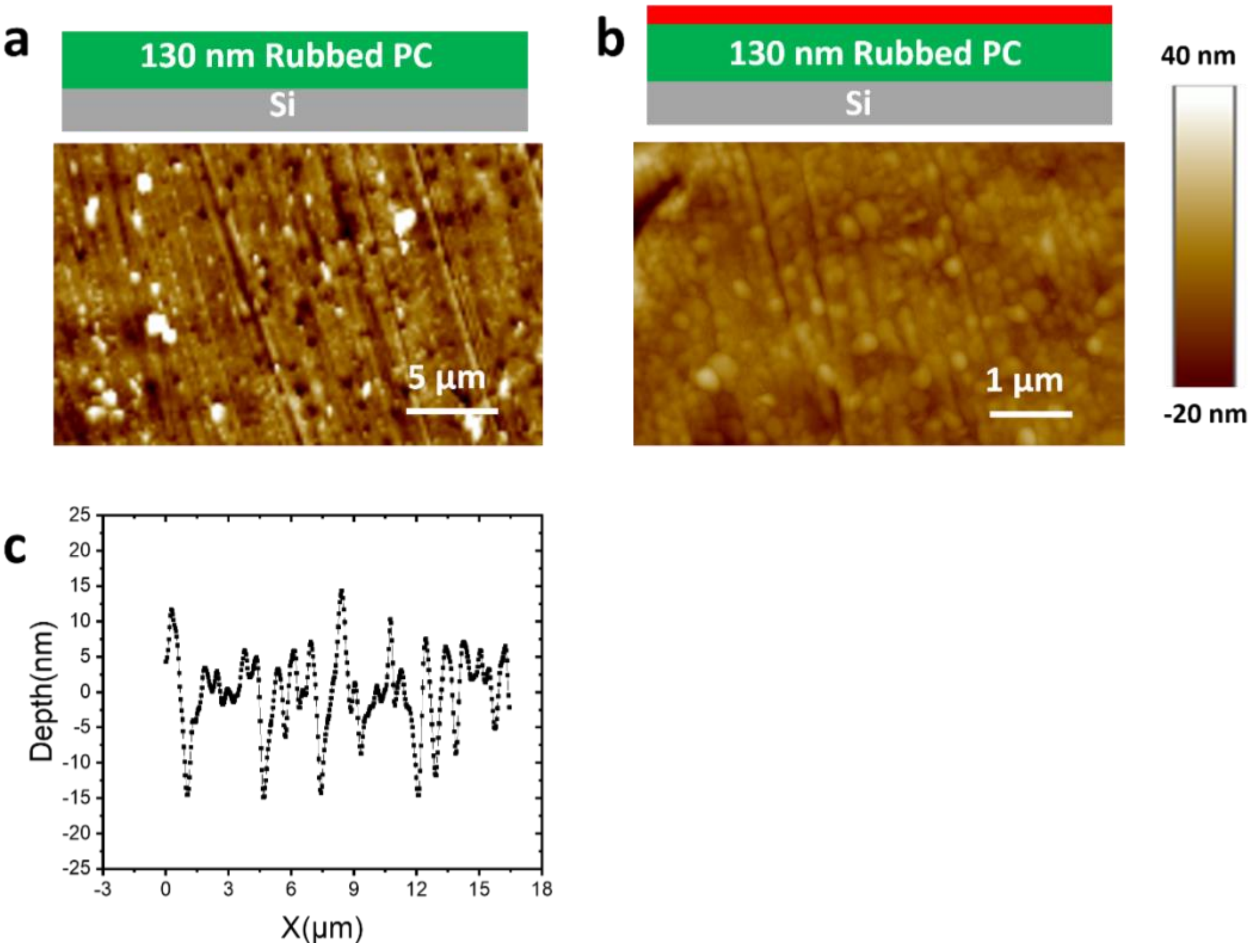

*Figure S2 AFM images of the surface of polycarbonate layer (a) and template (b). (c) Depth as the function of distance, perpendicular to the rubbing direction, calculated from (a).*

**3. GIWAXS analysis: biaxial anisotropy of phen-ester**

Figure S3 shows GIWAXS of an anisotropic bilayer and an isotropic sample (on silicon wafer). Similar patterns can be observed, showing that the out-of-plane orientation is preserved under in-plane alignment. Comparing the bilayer and isotropic film at $T_{sub}$ = 386 K, a more ordered hexagonal structure can be observed in isotropic film (more concentrated pattern). This is possibly due to the roughness of the PC layer.

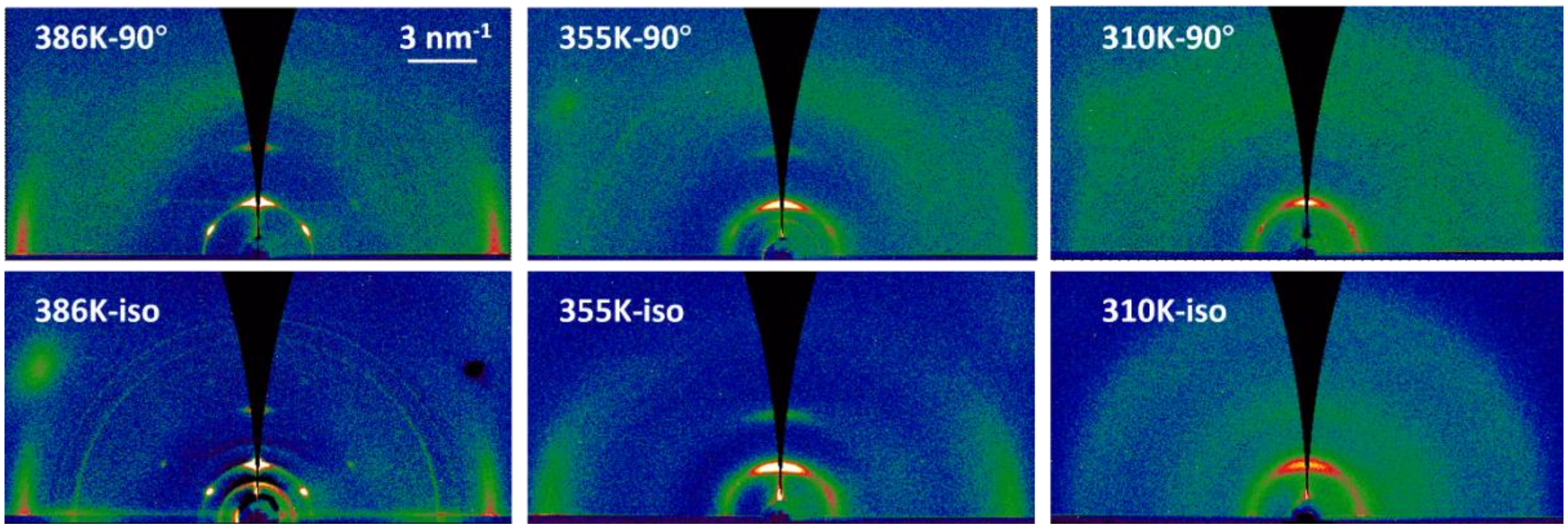

*Figure S3 GIWAXS of films on anisotropic bilayer (top) and isotropic (bottom) substrate for phen-ester and itraconazole. Substrate temperatures are indicated. Note that for all the GIWAXS shown in the session, the whole q range is same as in main text.*

Additional to the GIWAXS patterns shown in Figure 2, Figure S4 shows more diffraction patterns for phen-ester films at lower $T_{sub}$. No in-plane anisotropy can be observed for bilayers below 340 K and monolayers below 370 K. Notably, hexagonal peaks (peak-H) with weak intensity can still be detectable in 310 K bilayer. This is possibly from the diffraction of the underneath templating layer ($T_{sub}$ = 392 K), even though only with 30 nm thickness. It is also possible that the edge-on molecular orientation in the templating layer generates a week homeotropic alignment, which benefits the formation of hexagonal packing.

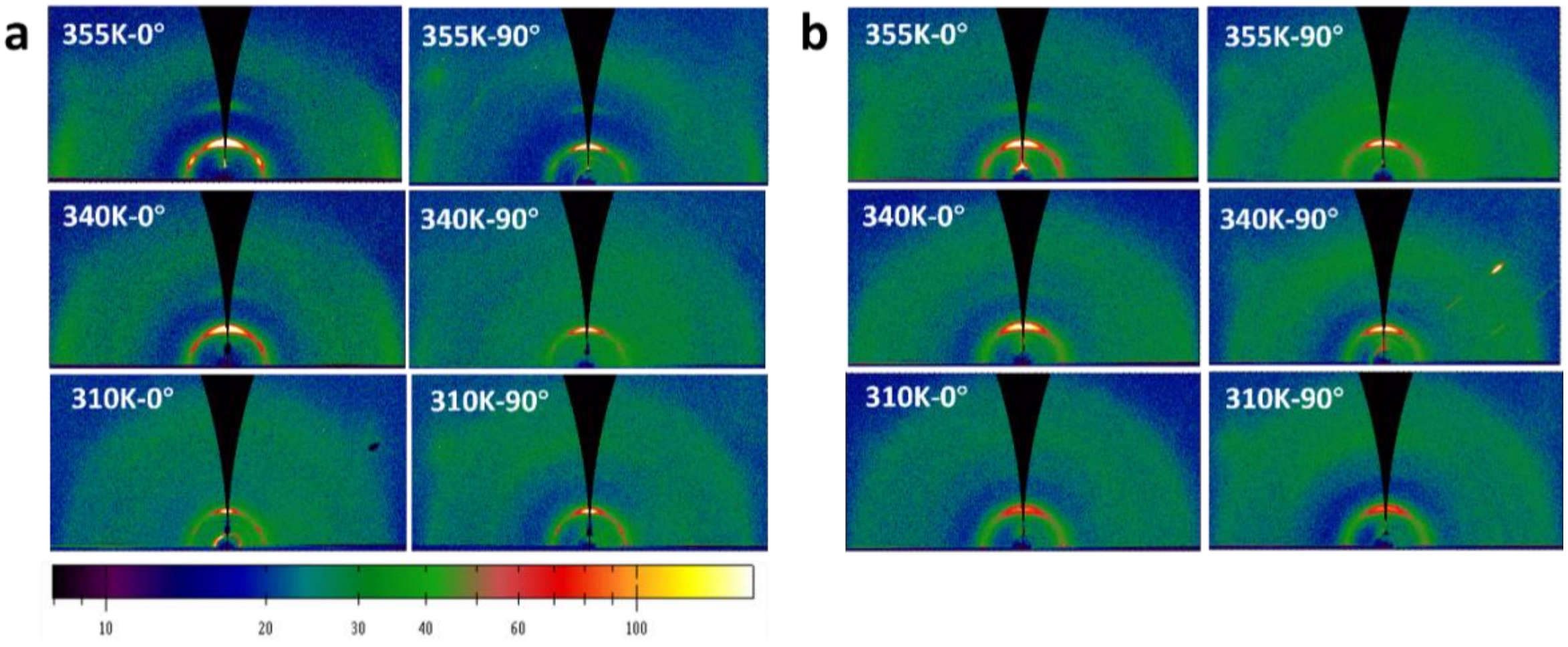

*Figure S4 GIWAXS patterns for (a) bilayer and (b) monolayer at lower $T_{sub}$ .*

An example of fitting to obtain peak areas $S_H$ and $S_\pi$ is shown in Figure S5(a), calculated from 386 K bilayer at $\psi_{view}$ = 90°. The diffraction of PC layer is fitted with two gaussian peaks are, indicated as PC-1 and PC-2 in Figure S5(a). $r(\psi_{view}) = S_H/S_\pi$ of 370 K monolayer is shown in Figure S5(b) (same as the inset of Figure 2(d)).

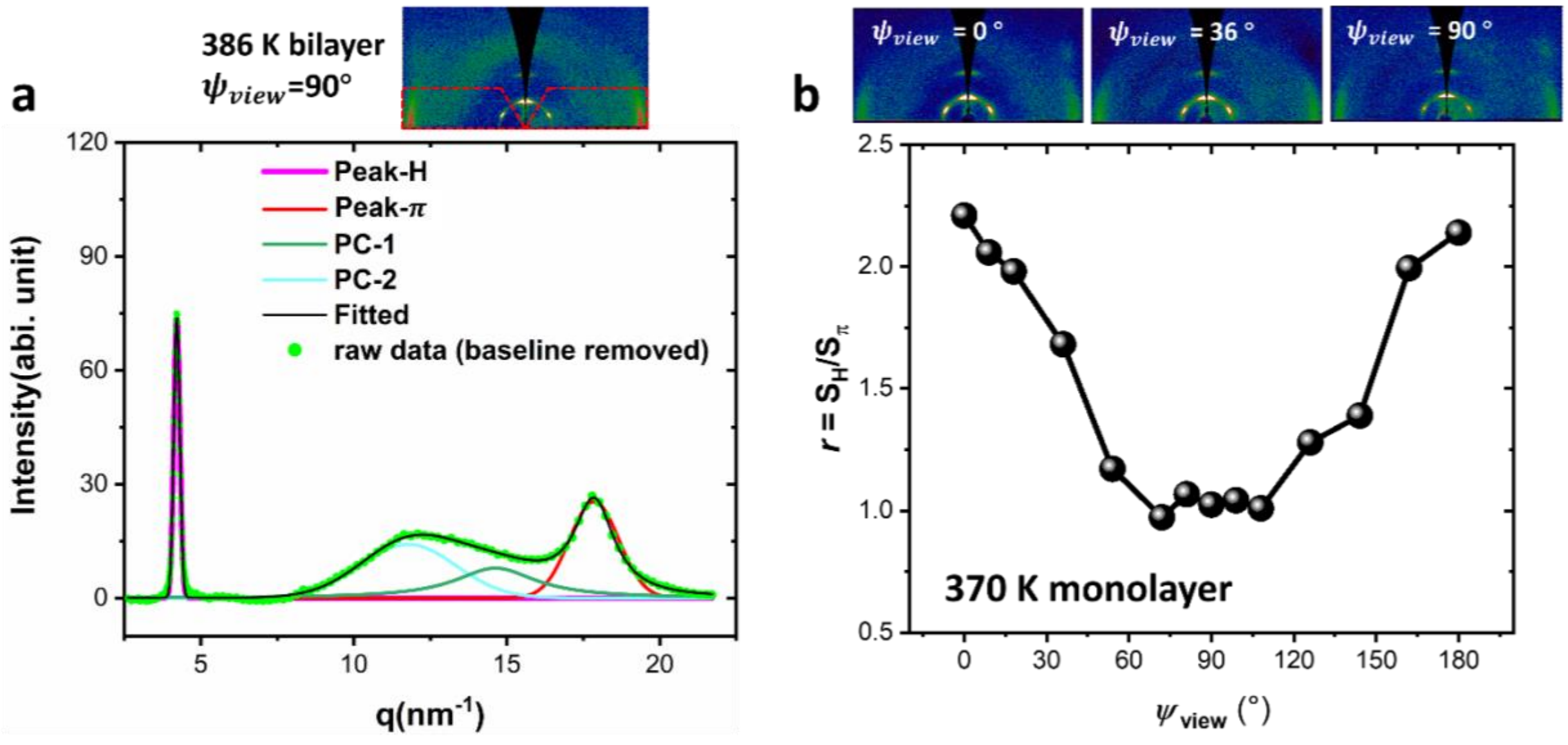

*Figure S5 (a) Fitting of 386 K bilayer at $\psi_{view}$ = 90 °. (b) $r(\psi_{view}) = S_H/S_\pi$ of 370 K monolayer.*

To quantitatively compare the in-plane anisotropy to more standard analysis [2], we calculated the orientation parameter $S_{2D}$, defined as [2]:

$$S_{2D} = 2\langle cos^2\psi\rangle - 1$$

, where

$$\langle cos^2\psi\rangle = \frac{\int_{0°}^{90°} I(\psi)cos^2\psi d\psi}{\int_{0°}^{90°} I(\psi)d\psi}$$

In this way, a higher $S_{2D}$ value means that more hexagonal columns are aligned along the direction of the X-ray beam.

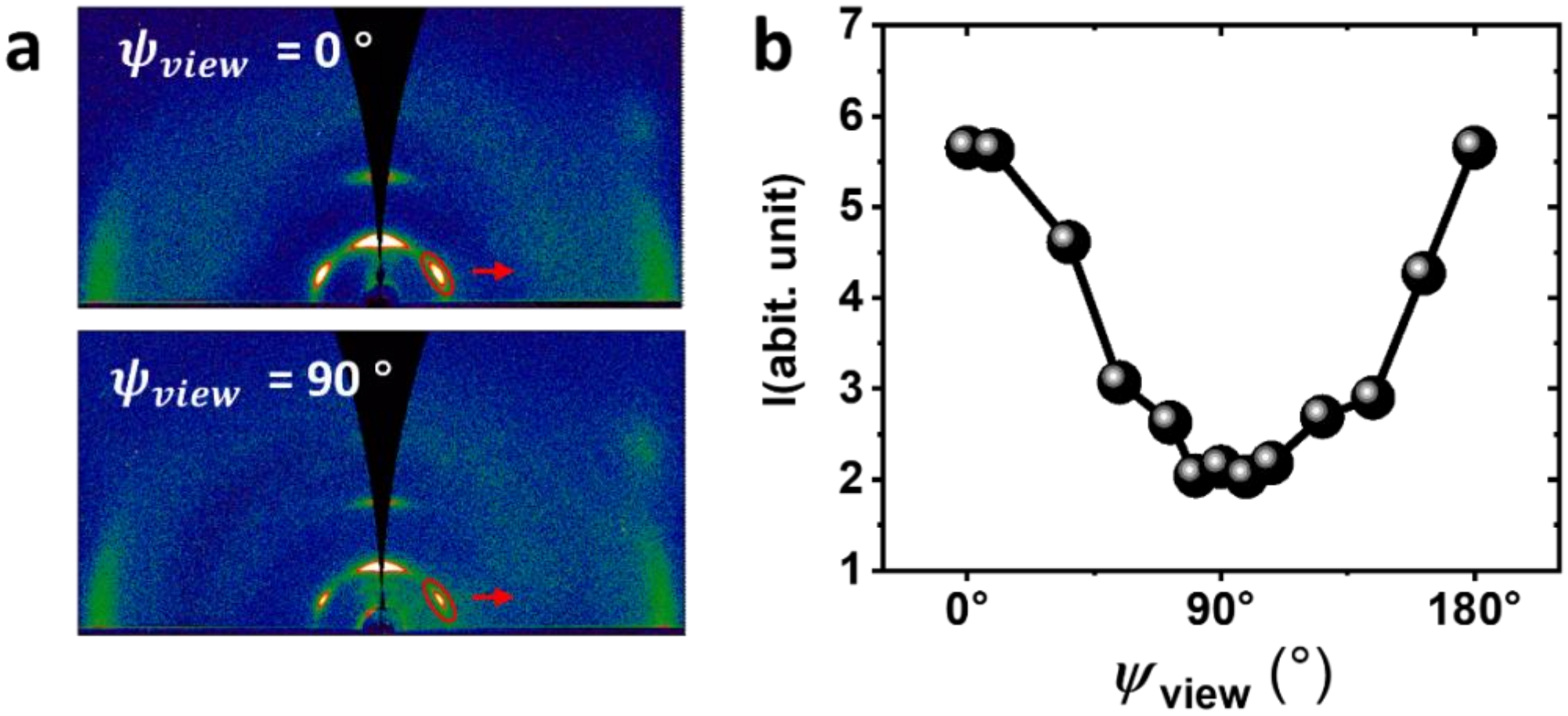

*Figure S6 Results for 370 K monolayer: (a) GIWAXS patterns at $\psi_{view}$ of 0 ° and 90 °, respectively. (b) Normalized intensity as the function of $\psi_{view}$.*

$I(\psi)$ is calculated in a small region in GIWAXS patterns, containing only peak-H at 60 ° (indicated in Figure S6(a), for 370 K monolayer). The intensity is normalized by the intensity at large $q$ around 25 $nm^{-1}$ after subtracting background intensity, and background intensity is defined as the average intensity at $q$ of 5.5 and 3.5 $nm^{-1}$. The normalized intensity $I(\psi)$ for the 370 K monolayer sample is plotted in Figure S6(b). Integration is performed after interpolation of data in Figure S6(b) and the calculated $S_{2D}$ is 0.26. Correspondingly, $K$ is 2.4 in Figure 2(d).

While we only acquired data at many $\psi_{view}$ values for a single sample (as shown in Figure S6 above), we can estimate $S_{2D}$ for other samples by only knowing $I(0°)$ and $I(90°)$, by assuming that all samples have a sinusoidal intensity distribution. The curve shown in Fig. 6(b) can be approximately expressed as a sinusoidal distribution:

$$I(\psi) = \frac{I_{max} - I_{min}}{2}\cos(2\psi) + \frac{I_{max} + I_{min}}{2} = \frac{1}{2I_{min}}\left[\left(\frac{I_{max}}{I_{min}} - 1\right)\cos(2\psi) + \frac{I_{max}}{I_{min}} + 1\right]$$
$$= \frac{1}{2I_{min}}\left(\frac{2I_{max}}{I_{min}}cos^2\psi + 2sin^2\psi\right)$$

In this way, $\langle cos^2\psi\rangle$ dependent only on $\frac{I_{max}}{I_{min}}$:

$$\langle cos^2\psi\rangle = \frac{\int_{0°}^{90°}\left(\frac{2I_{max}}{I_{min}}cos^4\psi + 2sin^2\psi cos^2\psi\right)d\psi}{\int_{0°}^{90°}\left(\frac{2I_{max}}{I_{min}}cos^2\psi + 2sin^2\psi\right)d\psi}$$

$$\langle cos^2\psi\rangle = \frac{\int_{0°}^{90°}\left((\frac{2I_{max}}{I_{min}}-2)cos^4\psi + 2cos^2\psi\right)d\psi}{\int_{0°}^{90°}\left((\frac{2I_{max}}{I_{min}}-2)cos^2\psi + 2\right)d\psi}$$

Since we know

$$\int_{0°}^{90°} cos^4\psi d\psi = \frac{3\pi}{16}$$

$$\int_{0°}^{90°} cos^2\psi d\psi = \frac{\pi}{4}$$

, we have

$$\langle cos^2\psi\rangle = \frac{\frac{3\pi}{8}\left(\frac{I_{max}}{I_{min}}-1\right)+\frac{\pi}{2}}{\frac{\pi}{2}\left(\frac{I_{max}}{I_{min}}-1\right)+\pi} = \frac{\frac{3I_{max}}{I_{min}}+1}{\frac{4I_{max}}{I_{min}}+4}$$

This approach can be tested using the data for the 370 K monolayer, where our approximate method (using only $\psi_{view}$ values at 0° and 90°) yields an estimated $S_{2D} = 0.24$, quite close to the value obtained calculated from the entire set of $\psi_{view}$ values ($S_{2D} = 0.26$). For the sample with the highest orientation, the bilayer at $T_{sub}$ = 386 K, we measured $\frac{I_{max}}{I_{min}} \approx 6.7$, and we used this approximate method to obtain an estimate of $S_{2D} = 0.37$.

**4. GIWAXS of 380 K bilayer phen-ester on $Alq_3$ substrate**

GIWAXS patterns for the same film of 380 K bilayer in Figure 3 are provided here in Figure S7. Similar to observation on PC substrate, Peak-H and Peak-$\pi$ shows stronger intensity at 0 ° and 90 °, respectively.

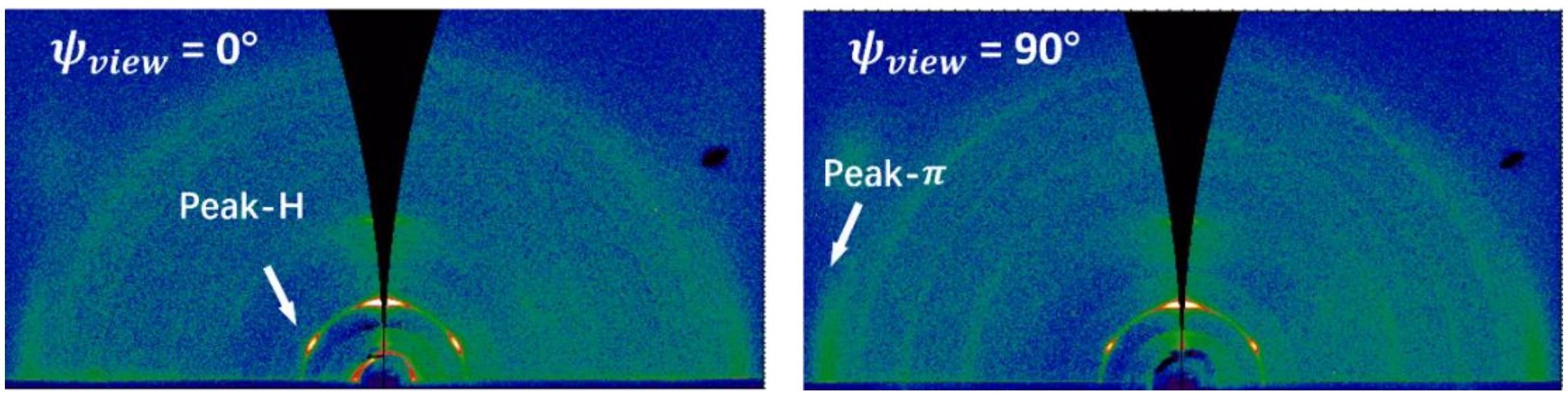

*Figure S7 GIWAXS patterns of 380 K bilayer on $Alq_3$ substrate, with $\psi_{view}$ = 0 ° and 90 °, respectively.*

## 5. GIWAXS of rubbed PC layer

The GIWAXS patterns of rubbed PC are shown in Figure S8. Stronger intensity can be observed with $\psi_{view}$ = 0°. In both patterns, the structure is slightly anisotropic, with layered structure at $q$ = 12.2 $nm^{-1}$.

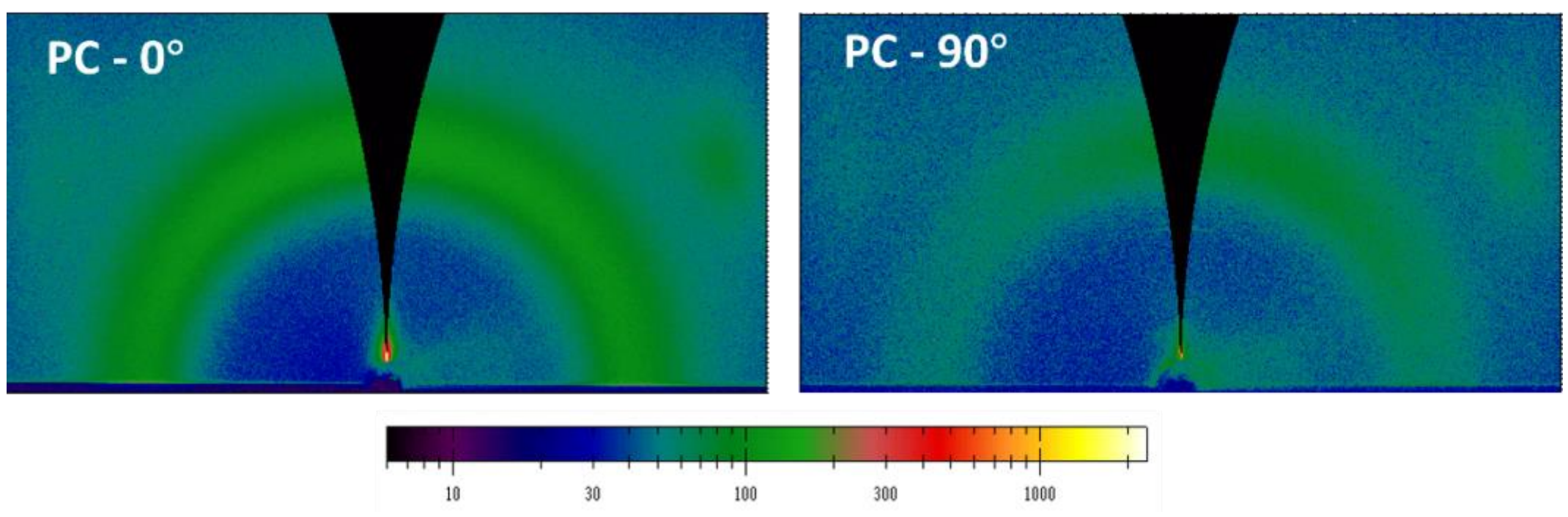

*Figure S8 GIWAXS patterns of rubbed PC with $\psi_{view}$ = 0 ° and 90 °, respectively.*

## 6. In-situ heating with GIWAXS

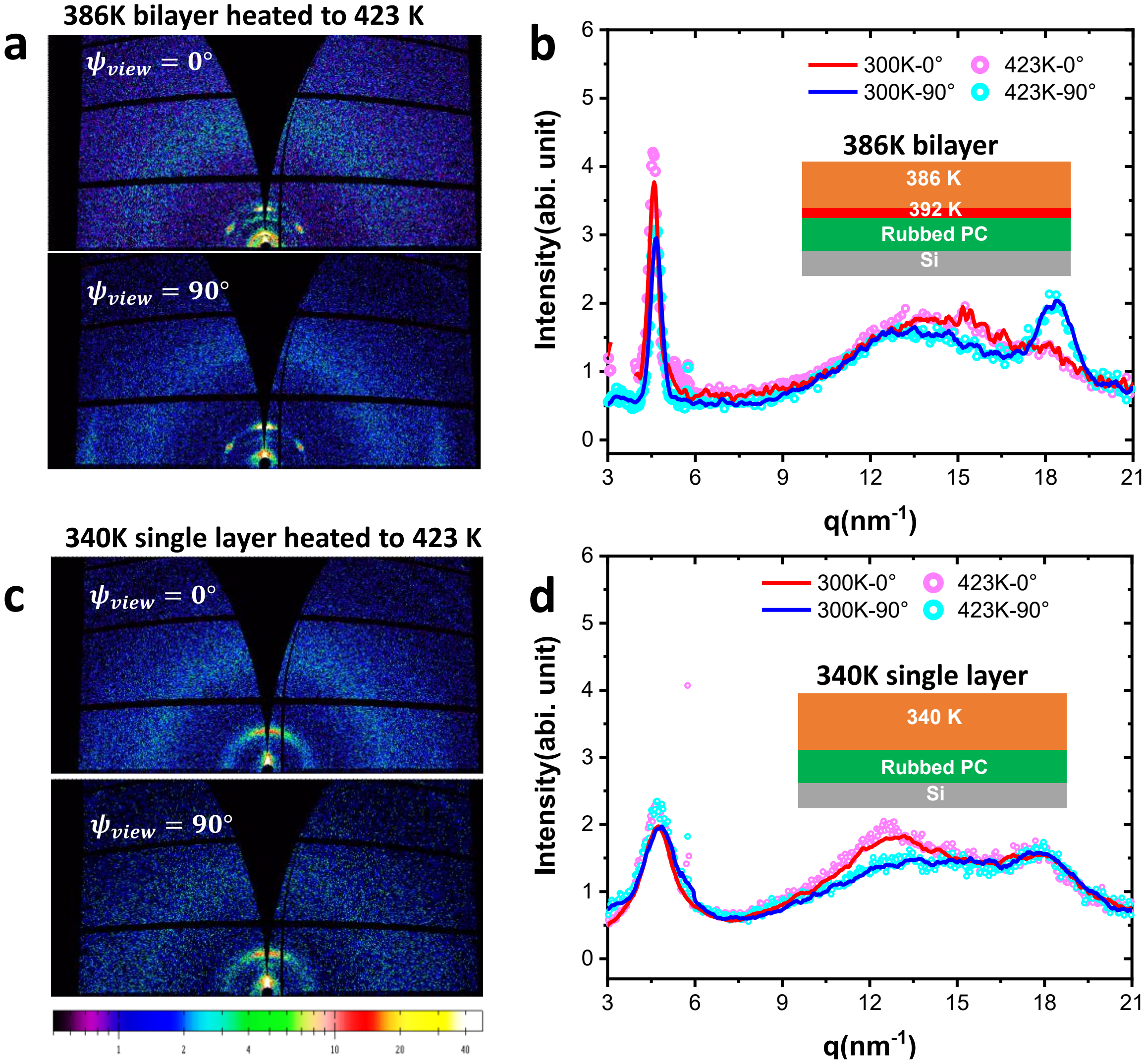

*Figure S9 (a) GIWAXS patterns of 386 K bilayer heated to 423 K. (b) 1D intensity of 386 K bilayer during heating from integration same as in Figure 2(c). (c) GIWAXS patterns of 340 K single layer heated to 423 K. (d) 1D intensity of 340 K single layer during heating.*

As mesogens can also be aligned in the fluid mesophase above $T_g$, it is interesting to directly compare this with the alignment effect by PVD. For this purpose, PVD films were heated from room temperature to 423 K (30 K over $T_g$, where phen-ester mesogens at equilibrium are in the columnar liquid crystal phase [3, 4]) with heating rate of 2 K/min. Structural information during heating was measured with *in situ* GIWAXS (Xeuss 2.0 system of Xenocs), at $\psi_{view}$ = 0 ° and 90 °, respectively. The frame rate was 300 s/image, so each image is an average over a 10 K temperature window.

Figure S9 shows that alignment in the equilibrium liquid crystal state at temperatures up to $T_g$ + 30 K is very slow. As shown in Figure S9(a), for a 386 K bilayer sample, GIWAXS patterns are almost identical before and after heating to 423 K and thus this sample fully retains the alignment provided by PVD. In Figure S9(c), for a 340 K single layer sample, in-plane hexagonal packing does not develop even after heating to 423 K, as the film remains nearly isotropic. These conclusions are confirmed by 1D intensity curves in Figures S8(b) and 3(d) (integrated in the same region as in Figure 2(c)). Thus for phen-ester mesogens in their liquid crystalline phase at 423 K, the mobility in a thin film does not suffice to alter or exceed the alignment achieved by PVD. On the other hand, aligned columnar packing can be achieved at $T_{sub}$ = 355 K by PVD employing a template layer (Figure S4). These observations show that PVD can achieve in-plane alignment of columnar packing at much lower temperatures than traditional templating with bulk liquid crystal phases.

GIWAXS with in-situ heating were measured in the University of Southern Mississippi with Xeuss 2.0 system of Xenocs, with wavelength of 0.154 nm. The distance between the sample and detector was 148 mm, calibrated with Silver behenate (AgBh). The samples were heated on a Linkam stage and measured in vacuum, with heating rate of 2 K/min and collection time of 300 s per image. This results in an averaging window of 10 K for each diffraction. The samples for observation at different $\psi_{view}$ are prepared in one single deposition and heated separately. Since the alignment ability can be lost/suppressed over $T_g$ of PC around 423 K, higher temperature was avoided.